\documentclass[10pt]{article}
\usepackage{graphicx}
\usepackage{amsmath}

\title{Self-Compression of High-Intensity Femtosecond
Laser Pulses in Low Dispersion Regime}

\bigskip
\author{I. G. Koprinkov$^1$, M. D. Todorov$^2$, M. E. Todorova$^3$, and T. P. Todorov$^1$                               \\
\small\it $^1$Department of Applied Physics, Technical University of Sofia, 1000 Sofia, Bulgaria,\\[-1.mm]
\small\it $^2$Faculty of Applied Mathematics and Informatics, Technical University of Sofia, 1000 Sofia, Bulgaria\\[-1.mm]
\small\it $^3$Integrated Technical College, Technical University of Sofia, 1000 Sofia, Bulgaria}
\date{}

\begin{document}

\maketitle
\bigskip

\begin{abstract}

Self-compression of femtosecond laser pulses and more than an order of magnitude increase of the peak intensity is found in a positive dispersion medium in low dispersion regime based on the (3+1)-dimensional nonlinear Schr{\"o}dinger equation. A method of high-intensity femtosecond pulse formation can be developed on that basis.

\end{abstract}

The propagation of high-intensity femtosecond optical pulses in bulk nonlinear medium has attracted considerable attention since their generation became a mature technology. The rigorous analysis of the pulse propagation is based on the (3+1)-dimensional nonlinear Schr{\"o}dinger equation (NLSE), in which diffraction, group velocity dispersion (GVD) of second order, and Kerr nonlinearity of third order form a basic set of optical processes \cite{1,2,3,4} capable to describe a broad range of realistic conditions. Additional terms in the NLSE are required to describe some particular features of the pulse propagation and ionization \cite{5,6,7,8,9}. The NLSE predicts different pulse behavior depending on the physical conditions, but some general features can be outlined. Thus, at negative GVD, simultaneous collapse of the pulse in space and time and formation of a spatiotemporal soliton is predicted1. In the more common case of positive GVD, however, the theoretical studies predict time broadening and splitting of the pulse \cite{2,3,4,5,6,7}. In a single study, initial pulse shortening is found while evidences from the conducted experiments are not shown \cite{4}. A time shortening at self-focusing is known for nanosecond pulses \cite{10}, where, however, the GVD plays a negligible role and has not been considered in the propagation equation \cite{11}. Such a nanosecond result cannot be directly extrapolated to the femtosecond pulses, for which the dispersion plays a crucial role. Self-compression (SC), {\it before splitting}, of high-intensity femtosecond laser pulses at positive GVD has been found experimentally in pure atomic and molecular gases \cite{12}. SC was also observed in air \cite{8}, highly ionized low pressure argon \cite{13}, and BK-7 glass in presence of ionization using negatively chirped pulses \cite{14}.

The pulse splitting is an important part of the femtosecond pulse dynamics and much attention was paid to its theoretical and experimental study \cite{2,3,4,5,6,7}. From the other side, while the SC is very interesting from a theoretical and experimental point of view, there is little knowledge about the set of processes and range of conditions at which SC may occur. Two cases can be well distinguished in the pulse propagation studies: relatively low-intensity pulses propagating in relatively high-dispersion media (fused silica, BK-7 glass) \cite{3,4,5,6}, and relatively high-intensity pulses (creating significant ionization) propagating in relatively low-dispersion media (low or high pressure gases) \cite{7,8,9,12,13}. In this Letter we focus on the not yet fully explored case of high-intensity femtosecond pulse propagation in a positive GVD medium in a low dispersion regime and negligible ionization. A minimal set of processes in the NLSE, sufficient to trigger the SC, is determined. More than two times of SC and more than an order of magnitude increase of the peak intensity are found.

The pulse propagation along $z$-direction including diffraction, GVD of second order and Kerr nonlinearity of third order is described by the NLSE for the complex field amplitude  ${\tilde E}(r,z,\tau)$ (in moving frame)
\begin{equation}
\label{eq1}
\frac{\partial {\tilde E}}{\partial z} - \frac{\rm i}{2 k} \nabla^2_{\perp} {\tilde E} + \frac{{\rm i} \beta_2}{2} \frac{\partial^2 {\tilde E}}{\partial \tau^2} - \frac{{\rm i} k n_2}{n_0} |{\tilde E}|^2 {\tilde E}=0,
\end{equation}
where standard notations are used \cite{4}. The field is normalized such that $I=|{\tilde E}|^2$  is the intensity. The nonlinear medium is pressurized argon of nonlinear refractive index of $n_2=9.8 \times 10^{-20}$ P cm$^2$/W, GVD of  $\beta_2=0.2$ P fs$^2$/cm, and $P$ is the pressure in atm. The initial pulse, tuned at 800 nm, is a linearly polarized chirp-free Gaussian in space and time of axial symmetry, ${\tilde E}(r,z=0,\tau)=E_0 \exp(-r^2/2r_0^2-\tau^2/2\tau_0^2)$, having 100 fs time duration (full width at half maximum (FWHM)) of the intensity profile and $W_0=0.3$ mJ energy.

The propagation equation (\ref{eq1}) does not include irreversible losses and non-instantaneous processes. Among these, the ionization has very important role as it strongly modifies the material parameters and thus, the pulse propagation, and, being a highly nonlinear process, is very difficult to be controlled. Avoiding the ionization of the medium can be achieved reducing the input pulse energy so that the peak intensity, increased by the self-focusing, does not reach the \begin{figure}[h]
\begin{center}
\includegraphics[width=2.2in]{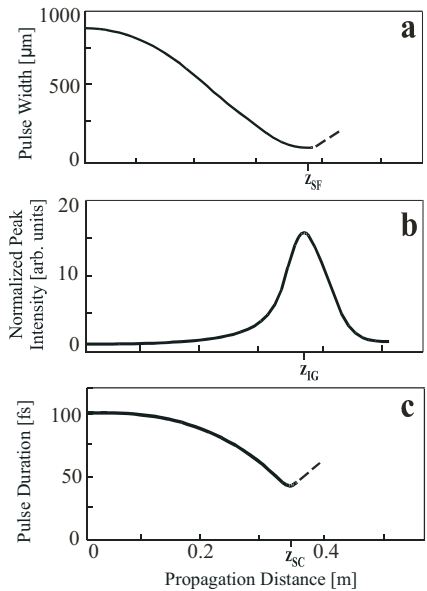}
\caption{Evolution of the transversal pulse width (FWHM) (a), peak intensity (b), and pulse duration (FWHM) (c) with the propagation distance. The dashed line indicates where the respective quantities are not well defined due to the pulse splitting. }
\end{center}
\end{figure}
ionization ``threshold" while the self-focusing is arrested by the diffraction and the temporal splitting. As our simulations show, at $W_0=0.3$ mJ and $P=15$ atm, the peak intensity does not exceed $10^{13}$ W/cm$^2$ throughout the propagation distance. This is below the ionization ``threshold" at which non-negligible ionization of Ar occurs\cite{9}. Also, according to the experiment \cite{15}, $10^{13}$ W/cm$^2$ is less than the appearance intensity at which single argon ion is detected for 8-photon ionization at 586 nm, and it holds even better for 11-photon nonresonant ionization at 800 nm, as in our case. Finally, the atomic medium does not exhibit in practice non-instantaneous properties \cite{5,6}. Thus, the propagation equation (\ref{eq1}) is appropriate to describe the behavior of the femtosecond pulses at the specified conditions. Such pulses can be routinely generated by the conventional oscillator-chirped pulse amplifier Titan:Sapphire laser systems.

The lack of substantial losses means the condition for conservation of the pulse energy $W(z)$ holds, {\it i.e.},
\begin{equation}
\label{eq2}
W(z)=\int\limits_{S_{\infty}} \int\limits_{-\infty}^{+\infty} |{\tilde E}(r,z,\tau)|^2 {\rm d}s {\rm d}\tau = W_0 ={\rm const},
\end{equation}
where the integration is taken over the transversal cross-section $s$ and the local time $\tau$ of the pulse. This resembles the normalization condition of the quantum mechanical wave function, while here Eq.(\ref{eq2}) plays role of normalization condition of the field amplitude. This allows to control the absolute value of the field strength and the pulse intensity.

The NLSE is solved numerically for the specified initial conditions by using a split-step method on a two-dimensional grid (in a polar coordinate system) of 200 transversal grid nodes and 100 temporal nodes. The evolution of the transversal pulse width $2r_0$  (FWHM at $\tau=0$  plane), the normalized peak intensity $|E_0 (r=0,z,\tau=0)/E_0(r=0,z=0,\tau=0)|^2$, and the pulse duration $\tau_0$ (FWHM) of the intensity profile versus the \begin{figure}[h]
\begin{center}
\includegraphics[width=2.5in]{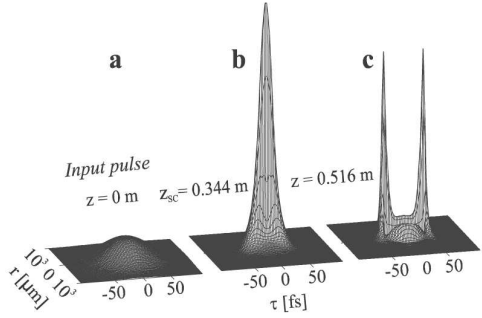}
\caption{The spatiotemporal shape of the input pulse (a), the compressed pulse at $z=z_{\rm SC}$ (b), and the split pulse at $z=0.516$ m (c). }
\end{center}
\end{figure}
propagation distance $z$ at $P=15$ atm of argon pressure is shown in Figures 1(a), (b), (c), respectively. The pulse rearrangement begins with a self-focusing. This confines the pulse in the transversal direction reaching minimal width at $z = z_{\rm SF}=0.378$ m, Figure 1(a). The self-focusing results in an increase of the peak intensity, Figure 1(b), and, what is nontrivial, SC of the pulse in time {\it before splitting}, Figure 1(c). The evolution of the spatiotemporal pulse shape is shown in Figure 2. As can be seen, a strongly compressed single pulse of increased intensity is formed. The space and time intensity profiles at the position of maximum SC are shown in Figures 3(a) and (b), respectively. The position of the maximum self-compression $z_{\rm SC}= 0.344$ m is close to the positions of maximal intensity gain, $z_{\rm IG}=0.367$ m, Figure 1. This is very important for the practical application as it allows taking \begin{figure}[h]
\begin{center}
\includegraphics[width=2.3in]{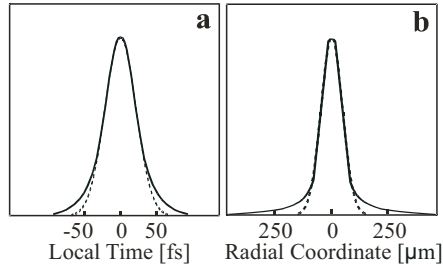}
\caption{Space $|E(r,z=z_{\rm SC},\tau=0)|^2$ (a) and time $|E(r=0,z=z_{\rm SC},\tau)|^2$ (b) intensity profiles of the pulse at maximum compression, $z=z_{\rm SC}$. The respective Gaussian shapes (dashed lines) are shown for comparison. }
\end{center}
\end{figure}
advantage from both the SC (more than 2 times at $z_{\rm SC}$) and the intensity gain (more than 15 times at $z_{\rm IG}$). If $z_{\rm SC}$ is at the end of the medium, a strongly compressed and amplified in intensity single pulse will be generated, which carries almost all of the input pulse energy. The propagation of the pulse beyond $z_{\rm IG}$ leads to the usual pulse splitting, Figure 2(c), which results in rapid fall of the peak intensity at $\tau=0$, Figure 1(b). The magnitude of the time compression and the intensity gain as well as the position of the strongest time ``focusing" $z_{\rm SC}$ and the intensity gain $z_{\rm IG}$, can be controlled by the pulse ({\it e.g.}, energy) and the medium (pressure) parameters.

A general explanation of the SC in this case (aside of the numerical simulations) can be done based on the energy conservation condition, Eq.(\ref{eq2}). This imposes conservation of the four-dimensional ``volume", as the pulse energy $W$ can be considered, which is distributed in the two transversal directions (the cross-section of the pulse), the longitudinal direction (the local time), and the ``vertical" direction (the intensity $|{\tilde E}|^2$). The self-focusing confines the pulse toward the longitudinal axis, $r=0$. The conservation of the energy/``volume" requires an expanding in the rest two directions, the intensity and the time duration ones. Due to the low dispersion of  the argon gas, the pulse energy does not substantially spread in time (along a given distance), while the strong increase of the pulse intensity (more than 15 times, Figure 1(b)) results in an effective, but {\it real}, shortening of the pulse, based on the FWHM characterization. To verify the importance of the GVD for the SC, the pulse propagation has been simulated in a hypothetical medium of increased GVD, keeping the other parameters constant. Thus, if GVD exceeds $1.95$ fs$^2$/cm, no SC takes place while continuous time broadening and splitting the pulse is observed, because the increased GVD speeds up the time expansion and the splitting of the pulse and SC cannot be achieved. To the best of our knowledge, such a concept of femtosecond pulse compression is formulated for the first time and will be called {\it self-compression in low dispersion regime}. It is based on a geometrical rearrangement of a pulse of almost constant energy. An experimental method of pulse compression and intensity gain can be developed providing the ionization (and other losses) is negligible. It can be scaled up to much higher intensity/energy pulses using medium of higher ionization potential, {\it e.g.}, Ne or He. Keldysh adiabatic parameter \cite{15} can be used as a scaling parameter providing similar ionization condition as in Ar.

The general propagation properties of the pulse up to the stage of maximum SC in these simulations well matches the experimentally observed behavior in pressurized argon, see Figure 1(b) in Ref.\cite{12}  and Figure 2 in Ref.\cite{16}. From the other side, the higher pulse energy used in these experiments results in peak intensity at maximum SC as high as $5\times 10^{13}$ W/cm$^2$ or more, and the ionization and above-cubic nonlinear effects are no more negligible \cite{9}. The experimentally observed stabilization of the pulse shape, once the SC is achieved, can be attributed namely to such additional processes \cite{12,16}.

In conclusion, self-compression of high intensity femtosecond pulses is found as a result of the interplay of the diffraction, group velocity dispersion of second order and Kerr nonlinearity of third order in a low dispersion regime. These three processes are sufficient to trigger the self-compression in a positive GVD medium in low dispersion, while additional mechanism is required for a stable propagation of the compressed pulse. A method of pulse compression and intensity gain of high-intensity femtosecond laser pulses below the ionization ``threshold" can be developed on that basis.

\end{document}